# Novel Regime of Operation for Superconducting Quantum Interference Filters

A.V. Shadrin[a], K.Y. Constantinian[a], G.A. Ovsyannikov[a], S.V. Shitov[a], I.I. Soloviev[b], V.K. Kornev[b], J. Mygind[c].

[a] *Kotel'nikov Institute of Radio Engineering and Electronics Russian Academy of Sciences 125009 Moscow, Russia.*

[b] *Moscow State University, Physics Department, 119992 Moscow, Russia.*

[c] *Technical University of Denmark, Institute of Physics, DK-2800 Kgs. Lyngby, Denmark.*

**A new operating regime of the Superconducting Quantum Interference Filter (SQIF) is investigated. The voltage to magnetic field response function, *V(H)*, is determined by a Fraunhofer dependence of the critical current and magnetic flux focusing effect in Josephson junctions (F-mode). For SQIF-arrays made of high-Tc superconducting bicrystal Josephson junctions the F-mode plays a predominant role in the voltage-field response *V(H)*. The relatively large superconducting loops of the SQIF are used for inductive coupling to the external input circuit. It is shown that the output noise of a SQIF-array measured with a cooled amplifier in the 1–2 GHz range is determined by the slope of the *V(H)* characteristic. Power gain and saturation power were evaluated using low frequency SQIF parameters. Finally, we consider the influence of the spread in the parameters of Josephson junctions in the SQIF-array on the *V(H)* characteristic of the whole structure.**

Microwave amplifiers based on superconducting quantum interference effect in two parallel connected Josephson junctions (SQUID) are characterized by a noise temperature close to the quantum limit (see, e.g., [1, 2]). The SQUID-amplifiers as other Josephson devices without feedback possess a low saturation power, which is proportional to the characteristic voltage $V_0 = I_C R_N$ where $I_C$ is the critical current and $R_N$ is the normal state resistance of the Josephson junction (JJ). $V_0$ is 200÷300μV at temperature T = 4.2K for JJs made from low-$T_C$ superconductors (LT$_c$S) [2], and reaches $V_0$ = 1mV at T = 77 K for bicrystal JJs made from high-$T_C$ superconductors (HT$_c$S) [3].

Using an array of JJs can increase the saturation power. A series-coupled LT$_c$S SQUID array amplifier reported in [4] demonstrated an increase of the output signal proportional to the number N of SQUIDs. The spread in $I_C$ and $R_N$ of the JJs significantly restricts the application of series-connected JJ or SQUID arrays.

Recently, superconducting quantum interference filters (SQIF-array) [5–7] forming arrays of series-connected SQUIDs with sophistical distribution of SQIF-loop areas in order to have a single peak in the *V(H)* curve was suggested and studied for microwave amplification [8, 9]. However, an application of incoherent interference from different SQIF–loops is also possible. Large inductances of the SQIF-loops $L_i >> L_J = \Phi_0/2\pi I_c$ ($\Phi_0$ is magnetic flux quantum) provide better coupling with the external circuits but suppress the main peak in the *V(H)* function, and in turn eliminates the possibility of using usual SQUID mode operation (S-mode).

Also very sensitive *V(H)* magnetometers based on one-dimensional series arrays of HT$_c$S bicrystal JJ has been reported [10]. Here the Fraunhofer dependence of critical current for JJ $I_c(H)=I_c(0)\sin(\Phi_J/\Phi_0)$, (where $\Phi_J=\pi\mu_0 H a_J$ is magnetic flux in a JJ with effective area $a_J$) and flux focusing effect due to geometry of the bicrystal JJ were used (we call this operation regime F-mode). A voltage transfer function *dV/dH* as large as 7500 V/T was obtained with 105 junctions in an array that is similar to the LT$_c$S SQUID array [4].

In this paper we present results of an experimental investigation of series-connected SQIF-array made from HT$_c$S bicrystal JJs operating at f = 1-2GHz in the F-mode. Unlike usual series-connected JJs arrays both circulating currents and induced magnetic flux contribute to the amplification process in our SQIF-arrays. The current in the SQIF-loops induced by an external magnetic field is – so to say - transformed to a current circulating in the JJ. This provides the good coupling to the external circuit.

Bicrystal NdGaO$_3$ (NGO) substrates were used for samples fabrication. NGO is characterized by a tolerable permittivity ($\varepsilon$ = 25) and low microwave losses (tg$\delta$<$10^{-3}$). The devices were patterned using ion-plasma and chemical etching of YBa$_2$Cu$_3$O$_x$ (YBCO) film deposited by dc sputtering at high oxygen pressure. Fabrication details for the bicrystal JJ have been described elsewhere [3]. For comparison SQUIDs and series-connected SQUID- arrays [8] were fabricated. Fig.1 shows the topology of a SQIF-array designed for use as microwave amplifier: the input line (Au film) was deposited over the SiO$_2$ insulator layer, the bottom layer is the YBCO film which forms the SQIF-array located inside the input line, which provides the RF coupling of the input signal (see Fig.1, bottom inset). The output signal is taken directly from the SQIF-array.

We measured *I–V*-curves and magnetic field-to-voltage response *V(H)* at low frequency as well as the output noise power $P_N = k_B T_N \Delta f$ in the frequency band $\Delta f$ = 1GHz. For the latter we used a cryogenic preamplifier with noise temperature $T_{A1}$ = (8±2)K and gain G = 21dB followed by a room temperature amplifier with $T_{A2}$ = 130K and G = 40dB. The output signal was simultaneously recorded on a spectrum analyzer and detected by a quadratic detector built into the room amplifier.

Fig. 2 shows a family of $V(H)$ characteristics plotted at different dc bias currents, $I_b$. The experimental $V(H)$ curves clearly show the single F-mode dip around $H = 0$ with small lateral voltage modulation. The both the reference SQUID and the series-connected SQUID-array showed a pronounced superposition of S- and F-mode resulting in a strong lateral modulation [8]. The residual lateral modulation of the SQIF-array was additionally suppressed by increasing the bias current $I_b$, leading also to a smoothing of the wings of the main $V(H)$ dip. The expected increase in the height of the $V(H)$ response, being proportional to $V_S = \Sigma V_{0i}$, which increases with the number N of SQIF-loop each contributing with $V_{0i} = I_{Ci}R_{Ni}$. However, the experimental $V_S$= 8mV was smaller than the estimated $\Sigma V_{0i}$= 20mV for a SQIF-array with N = 20. A numerical simulation of the $V(H)$ dependence made using the PSCAN program [11] is shown in Fig.3. One can see that for SQIF-arrays with N > 10 (see Fig.3) the S-mode modulation is suppressed. Note, the experimental width of the F-mode dip is well fitted using an effective JJ area $S_{eff} = 30\mu m^2$ which is considerably larger than the area $w\lambda_L$= 1.5$\mu m^2$ where $w$ is the JJ width and $\lambda_L$ = 0.15$\mu$m is the London penetration depth in YBCO. This is in agreement with a strong flux focusing effect in the bicrystal JJ [12] with the magnetic field applied perpendicular to the film. The external magnetic flux, produced by the input signal coil, induces screening current in superconducting loops of each of the SQIF elements. Since the width of the printed loops (equal to the geometrical width of the bicrystal JJ) are larger than $\lambda_L$ the current near the outer edge of the printed loop flows in opposite direction of the current in the inner edge. Finally, the current in the loop induced by the external magnetic field is transformed to the current circulating in the JJ as a differential current.

From the experimental magnetic diffraction pattern, $I_C(H)$, of the reference SQUID with loop area $S = 35\mu m^2$ we estimate a loop inductance $L = (15\pm5)$pH [13], which is much larger than $L = 1.25\mu_0 w$. The estimated spread in critical current of JJ in SQUID is $\delta I/I_C = (30\pm10)\%$[1]. For a mean value of $I_C = 100\mu A$ for JJs with width $w = 10\mu m$ we derive the normalized inductances in the SQIF-array to be in the range $l_i = L_i/L_J = 4.5$-91. Taking into account a 30% spread of the critical currents $I_{Ci}$ of JJ in SQIF-array a digital simulation gave a decrease of the F-mode dip and a total smearing out of the S-mode modulation as observed in the experiments. Besides, the range of $dV/dH$ linearity, determining the saturation power of the SQIF-array, is ten times larger than in the reference SQUID-array. For the SQIF-array operating in F-mode we obtain $dV/d\Phi$ = 40 mV/$\Phi_0$ (and $dV/dH$ = 270 V/T), while for SQUID in the S-mode we have $dV/d\Phi$ = 1 mV/$\Phi_0$ [8].

---

[1] Asymmetry of the JJs parameters in a SQUID for small inductance 0.2<$l$<2 yields an increase of the conversion factor $dV/dH$ compared to the dependence calculated for a symmetric SQUID [15]. If $l >> 1$ the $dV/dH$ approaches an asymptotic slope symmetric SQUID and decreases significantly.

The spectral density of the output noise temperature $T_N(H)$ measured in the frequency range f = 1–2 GHz for a SQIF-array is presented in Fig.4. The residual S-mode peaks of the $T_N(H)$ dependence are correlated with the $dV/dH$ function. The output power is up to 15 dB above the background noise level. The measured output noise signal from the SQIF-array can be interpreted as an incoherent superposition of voltage thermal fluctuations of JJs with a spectral density of $S_V = \Sigma(8kTR_d^2/R_N)[1+1/2(I/I_C)^2]$ with a flux-to-voltage conversion noise $S_\Phi = \Sigma(2kTL_n^2/R_N)(dV_n/d\Phi_n)^2$ [14]. For a bias current $I_b$ = 1.25mA and $R_d$ = 30 Ω at the $T_N(H)$ peak we get $S_\Phi \gg S_V$. It indicates that the observed output signal is dominated by the flux-to-voltage conversion $(\Sigma L_i dV_i/dH_i)^2$. For the experimental conditions $dV/d\Phi$=40 mV/$\Phi_0$ the SQIF-array is characterized by a maximal flux sensitivity $\delta\Phi = 10^{-4} \Phi_0$. The lateral deviations of the $T_N(H)$ waveform and the corresponding $V_\Phi(H)$ dependence could be caused by residual contributions from the S-mode[2].

In order to estimate the power gain $G = \dfrac{M^2(dV/d\Phi)^2}{R_d R_s}$ [17] we take $R_s$ = 50Ω as the microwave source resistance, the dynamic resistance $R_d$ = 30 Ω, a mutual coupling inductance $M$ = 2.4 $10^{-11}$ H, assuming a geometrical circuits coupling coefficient α = 0.2 (for SQIF layout shown in Fig.1). Using the experimental value $dV/d\Phi$ = 2 $10^{13}$ sec$^{-1}$, as obtained from measurements at low frequencies, we get a reasonable value of G = 20 dB.

For estimation of the saturation power of the SQIF-array in the F-mode we measured the output signal at low frequencies f = 49Hz – 90kHz. The expected saturation power of non-coherently operating SQIF-loops increases as the square root of N, as $P_S \propto \sqrt{N}$ [5-7]. An analysis of the 1$^{st}$ and the 2$^d$ harmonics an applied 900 Hz signal shows a linear dependence over 60 dB with slow deviations of the 2$^d$ harmonic distortion at about 1% (see inset to Fig.4). Note, a similar harmonic distortion was observed for a LT$_c$S SQUID-amplifier operating in the S-mode [18].

Summarizing, we have fabricated and studied SQIF-arrays operating in the F-mode when the magnetic flux-to-voltage transfer factor is mostly determined by a Fraunhofer dependence of the critical current of the Josephson junctions. The flux-to-voltage conversion factor in the F-mode is apparently lower than the originally suggested S-mode at the single SQIF-array peak, nevertheless giving a power gain G > 1 and a significant increase in saturation power and dynamic range. Because of the large inductances of the SQUID- loops a spread in parameters of Josephson junctions doesn't play a decisive role in the F-mode, leading to a decrease in the contributions of the SQUIDs in the array to the total gain, but allowing at the same time a good signal amplification.

---

[2] Note, that differences in the critical currents in large inductive SQUIDs ($l \gg 1$) may also enhance the S-mode contribution [16] to the resulting output signal of the SQIF-array.

The authors thank I.V. Borisenko, Yu.V. Kislinski, A.V. Kolabukhov, P.V. Komissinskiy, I.M. Kotelynski, A.V. Sobolev and D. Winkler for fruitful discussions and help in experiments. The work was partially supported by Physical division of Russian Academy of Sciences, Scientific school NSh-5008.2008.2, FP6 European Union program NMP3-CT-2006-033191, RFBR 08-02-00487, ESF programs AQDJJ and THIOX and ISTC project 3743.

REFERENCES

1. M. Mück, Ch. Welzel, and J. Clarke, Appl. Phys. Lett., **82**, 3266 (2003).

2. G.V. Prokopenko, S. V. Shitov, I. L. Lapitskaya, et al., IEEE Trans. Appl. Supercond. **13**, 1042 (2003).

3. I.V. Borisenko, I.M. Kotelyanski, P. V. Komissinski et al., IEEE Trans. Appl. Supercond. **15**, 165 (2005).

4. M.E. Huber, P.A. Neil, R.G. Benson, et al., IEEE Trans. Appl. Supercond., **11**, 4048 (2001).

5. J. Oppenlander, T. Trauble, Ch. Haussler and N. Schopohl, IEEE Trans. Appl. Supercond., **11**, 1271 (2001).

6. V. Schultze, R. IJsselsteijn and H.-G. Meyer, Supercond. Sci. Technol. **19,** S411 (2006).

7. P. Caputo, J. Tomes, J. Oppenlander, et al., IEEE Trans. Appl. Supercond. **15**, 1044 (2005).

8. A.V. Shadrin, K.Y. Constantinian, and G.A. Ovsyannikov, Tech. Phys.Lett. **33**, 192 (2007).

9. O.V. Snigirev, M.L. Chukharkin, A.S. Kalabukhov et al, IEEE Trans. Appl. Supercond., **17**, 718 (2007).

10. S. Krey, O. Brugmann, and M. Schilling, Appl. Phys. Lett. **74**, 293 (1999).

11. V.K. Kornev, A.V. Arzumanov, Inst. Physics Conf. Ser., **158**, 627 (1997).

12. P.A. Rozenthal, M.R. Beasley, Appl. Phys. Lett., **59,** 26, 3482 (1991).

13. H. Hasegawa, Y. Tarutani, T. Fukazawa, and K. Takagi, IEEE Trans. Appl. Supercond., **8**, 26 (1998).

14. K.Enpuku, G.Tokida, T.Mruno, T. Minotani, J. Appl. Phys., **78,** 3498 (1995).

15. J. Muller, S. Weiss, R. Gross, R. Kleiner, D. Koelle, IEEE Trans. Appl. Supercond. **11,** 912 (2001).

16. G. Testa, E. Sarnelli, S. Pagano, C. R. Calidonna, and M. Mango Furnari J. Appl. Phys, **89**, 5145 (2001).

17. C. Hilbert, J.Clarke, J. Low Temp. Phys, **61**, 263 (1985).

18. M. Mück, J. Clarke, Appl. Phys. Lett. **78**, 3666 (2001).

**Fig.1.** Layout of SQIF-array for microwave amplifier consisting of 20 series-connected SQUID loops with areas in the range 35–700 µm². The width *w* of the Josephson junctions is 10 µm. The input line circuit consists of a top Au thin film (grey color) deposited over the SiO$_2$ buffer layer. The YBa$_2$Cu$_3$O$_x$ (YBCO) thin film is the bottom layer (dark color). The output circuit is a slot-line YBCO film with the SQIF-array located along the bicrystal boundary. The top inset shows a zoomed view of the bottom layer with a part of the SQIF array. Bottom inset shows a circuit with SQIF-array and pick-up loop.

**Fig.2** Magnetic field-to-voltage response, *V(H)*, of a SQIF array with 20 loops for different dc bias currents, $I_b$. The measurements were made at *T* = 4.2K. For the structure with critical current $I_C$ = 500µA the maximum in *V(H)* is $I_b$ = 1.1* $I_C$.

**Fig.3** Numerical simulation (PSCAN program) of *V(H)* for SQIF arrays with different number of loops. For the SQIF-array consisting of *N* = 20 loops the normalized inductances are in the range $l_i$ = 4.5 ÷ 91. F- and S-peaks are clearly seen for N>10.

**Fig.4** Magnetic field dependence of the equivalent noise temperature $T_N(H)$ for SQIF-array in the frequency band f = 1–2GHz (solid line). The dash-dot line shows the dV/dH(H) dependence. The bias current is $I_b$ = 1.25$I_C$, T = 4.2K. The insert shows the dependence of 1-st and 2-d harmonics of output signal $V_{out}$ vs input signal $V_{in}$ at frequency $f_{in}$ = 900Hz.

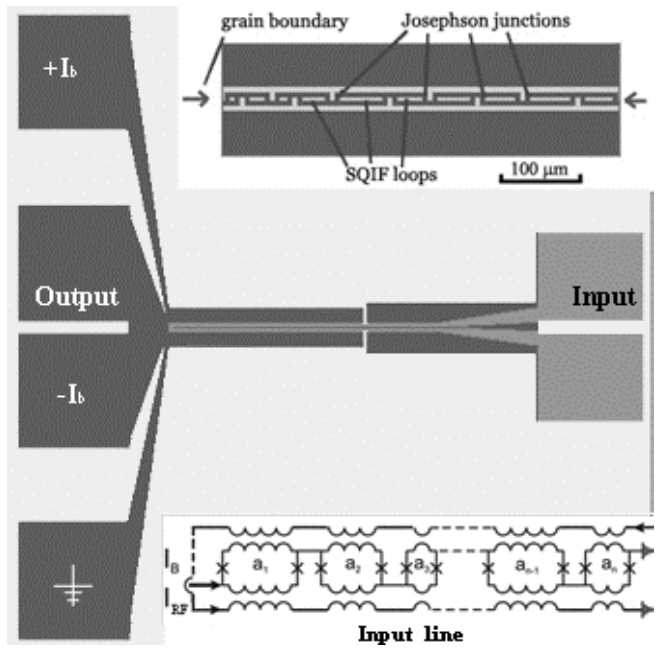

Fig. 1.

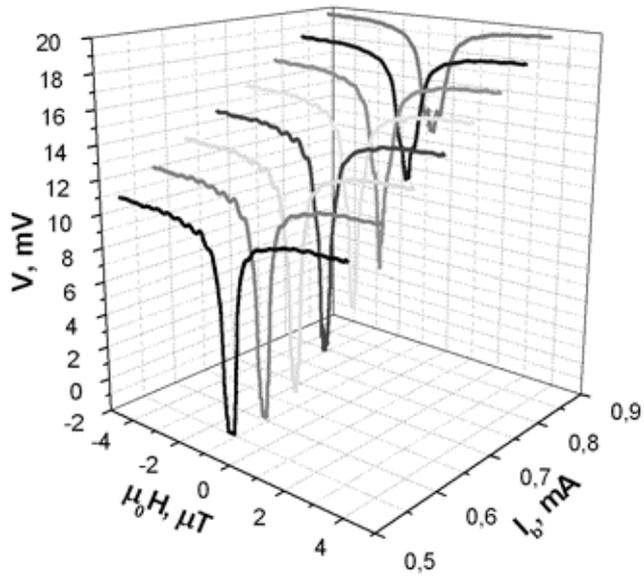

Fig. 2.

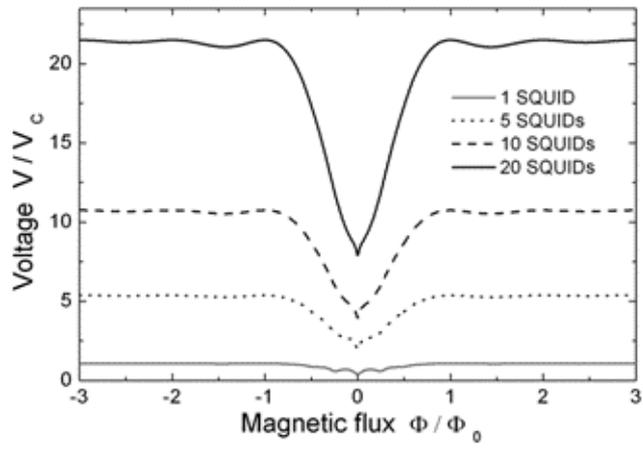

Fig. 3.

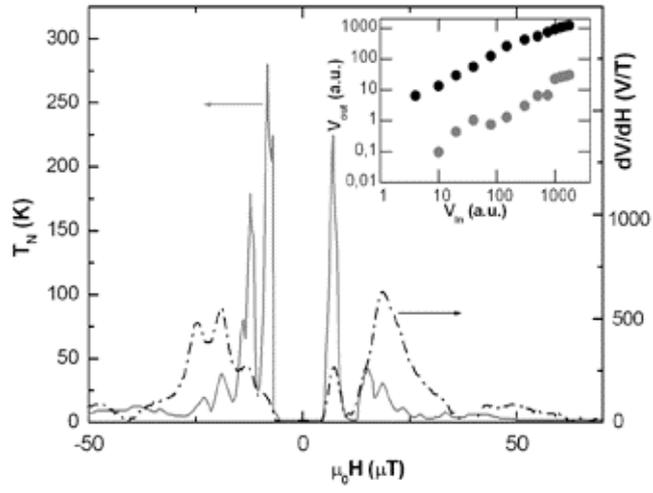

Fig. 4.